\documentclass[useAMS,usenatbib]{mn2e}
\usepackage{amsmath}
\usepackage{graphicx}
\usepackage{epstopdf}
\usepackage{amssymb,latexsym}
\usepackage{multicol}
\usepackage{color}

\begin{document}

\title[A search for open cluster Cepheids in the Galactic plane]{A
  search for open cluster Cepheids in the Galactic plane}
\author[Xiaodian Chen, Richard de Grijs and Licai Deng] {Xiaodian
  Chen$^{1,2,3}$, Richard de Grijs$^{1,2}$ and Licai
  Deng$^{3}$\\
$^{1}$Kavli Institute for Astronomy and Astrophysics, Peking
  University, Yi He Yuan Lu 5, Hai Dian District, Beijing 100871,
  China\\
$^{2}$Department of Astronomy, Peking University, Yi He Yuan Lu 5, Hai
  Dian District, Beijing 100871, China\\
$^{3}$Key Laboratory for Optical Astronomy, National Astronomical
  Observatories, Chinese Academy of Sciences, 20A Datun
  Road,\\ Chaoyang District, Beijing 100012, China}

\date{xxx}

\pagerange{\pageref{firstpage}--\pageref{lastpage}} \pubyear{2014}
\label{firstpage}

\maketitle

\begin{abstract}
We analyse all potential combinations of Galactic Cepheids and open
clusters (OCs) in the most up-to-date catalogues available. Isochrone
fitting and proper-motion calculation are applied to all potential
OC--Cepheid combinations. Five selection criteria are used to select
possible OC Cepheids: (i) the Cepheid of interest must be located
within 60 arcmin of the OC's centre; (ii) the Cepheid's proper motion
is located within the $1\sigma$ distribution of that of its host OC;
(iii) the Cepheid is located in the instability strip of its
postulated host OC; (iv) the Cepheid and OC distance moduli should
differ by less than 1 mag; and (v) the Cepheid and OC ages (and, where
available, their metallicities) should be comparable: $\Delta \log (t
\mbox{ yr}^{-1}) <0.3$. Nineteen possible OC Cepheids are found based
on our near-infrared (NIR) analysis; eight additional OC--Cepheid
associations may be genuine pairs for which we lack NIR data. Six of
the Cepheids analysed at NIR wavelengths are new, high-probability OC
Cepheids, since they lie on the near-infrared (NIR) period
($P$)--luminosity relation (PLR). These objects include TY Sct and CN
Sct in Dolidze 34, XX Sgr in Dolidze 52, CK Sct in NGC 6683, VY Car in
ASCC 61 and U Car in Feinstein 1. Two additional new OC Cepheids lack
NIR data: V0520 Cyg in NGC 6991 and CS Mon in Juchert 18. The NIR PLR
for our confirmed sample of OC Cepheids is $M_J= (-3.12 \pm 0.29)
\log(P \mbox{ day}^{-1})-(2.17 \pm 0.29)$ mag, which is in good
agreement with the best NIR PLR available for all Galactic Cepheids.
\end{abstract}

\begin{keywords}
methods: data analysis, stars: variables: Cepheids, open clusters and
associations: general, distance scale
\end{keywords}

\section{Introduction}

Cepheids are the brightest pulsating variable stars. They obey tight
period--luminosity relations (PLRs). The existence of such a PLR was
first found approximately a century ago based on photographic
observations of 25 Cepheids in the Small Magellanic Cloud
\citep{Leavitt1912}. For nearly one hundred years, many researchers
have been trying to establish and improve the relationship, mostly
using one of three common methods: (i) trigonometric parallaxes
\citep{Feast97, Benedict07}, (ii) Baade--Wesselink-type methods or,
equivalently, surface-brightness techniques \citep{Gieren97, Storm11}
and (iii) main-sequence or isochrone fitting. The latter method, when
applied to single-population star clusters, can be used as independent
constraint on the distances, ages and reddening values of Cepheids in
clusters, which thus enables its use as external calibrator of the
PLR. The key to the application of this method resides in selecting
high-confidence member stars of the relevant open clusters (OCs).

Since \citet{Irwin55} first matched the OC Cepheids S Nor and U Sgr
with NGC 6087 and M25, respectively, and following \citet{Feast57},
who first established stellar cluster membership based on
radial-velocity measurements, many researchers have contributed to
this field \citep[e.g.,][]{van den Bergh57, Efremov64, Tsarevsky66,
  Turner86, Turner10, Turner93, Baumgardt00, Hoyle03, An07,
  Majaess08}. To date, approximately 30 OC Cepheids have been found
and confirmed. This number is increasing as the available OC samples
are increasing. A good summary of the way in which calibration of the
Cepheid PLR is performed using the OC and Baade--Wesselink methods was
published by \citet{Tammann03}; current efforts focus on
high-precision membership research pertaining to OCs based on, e.g.,
data obtained as part of the VISTA Variables in the V{\'i}a L\'actea
(VVV) survey \citep{Majaess11, Majaess12a}. An eight-dimensional
search for bona fide Cepheids was undertaken by \citet{Anderson13},
based on a careful perusal of the existing literature. They found five
new OC Cepheids.

In this paper, our aim is to improve the completeness of the existing
body of OC Cepheids. We calculate proper motions using original data
rather than literature values for each cluster of interest, with as
main aim to find more OC--Cepheid associations with similar proper
motions. OC properties are also estimated independently based on
colour--magnitude-diagram (CMD) analysis and isochrone fitting so as
to improve the quality of the ages, reddening values and distances
contained in existing OC catalogues. In contrast to most previous
Cepheid PLR studies, we use near-infrared (NIR) observations to reduce
the impact of extinction, resulting in the determination of a reliable
$J$-band PLR.

In Section 2, we discuss our analysis of the OC and Cepheid catalogues
used in this study, as well as the five selection criteria applied to
find our OC Cepheids. The isochrone-fitting results, as well as
crucial reddening checks and our preliminary sample of possible OC
Cepheids, are covered in Section 3. We report and discuss the
confirmed, new, rejected and uncertain OC Cepheids, as well as the
issues affecting the completeness of our method, in Section 4. In
Section 5 we summarize our main conclusions.

\section{Method}

In this section we describe the data and method used to select
possible OC Cepheids. In general, we select OC Cepheid candidates by
comparing the Cepheids' positions with respect to the nearest OC
centres and assess the similarity of their proper motions, combined
with the Cepheids' loci in the cluster CMDs. Five criteria are used:
(i) a Cepheid must be located within 60 arcmin of an OC's centre; (ii)
the Cepheid's proper motion must be consistent with the $1\sigma$
proper-motion distribution of its host OC; (iii) the Cepheid lies in
the instability strip of its associated OC; (iv) the difference
between the Cepheid and OC distance moduli (DMs) $\Delta(m-M)_0 < 1.0$
mag; and (v) the difference between the Cepheid and OC ages, $\Delta
\log(t \mbox{ yr}^{-1}) <0.3$. In Sections 2.3 to 2.10, we outline the
details of our method.

\subsection{The open cluster catalogue}

This study is partially based on the \citet[][DAML02; version
  September 2013]{Dias02} OC catalogue. DAML02 contains approximately
2100 clusters, for which it includes entries related to the cluster
centre coordinates, size, proper motion and metallicity, if
measured. For each cluster, we compared the centre position with those
of the nearby Cepheids and obtained a sample of probable OC
members. For most OCs, we used the DAML02 OC centre and size to select
our initial samples of member stars. For some clusters characterized
by small apparent diameters ($< 6$ arcmin), we used the positions of
the bright Cepheids, combined with the cluster size, to select stars,
provided that the bright Cepheids are located close to the cluster
centres ($< 0.5$ arcmin). These small clusters are not very obvious
compared with the field stars, so that our use of the Cepheid
coordinates as a proxy for the cluster centres allows us to more
robustly select possible OC members. We also used the
WEBDA\footnote{http://www.univie.ac.at/webda/} database as
reference. It provides a large amount of information for OCs,
including their ages, distances, reddening values and CMDs in
different passbands.

Thus, we obtained the positions of our sample OCs from
DAML02. Photometric data pertaining to these OCs were obtained from
the homogeneous Two Micron All-Sky Survey (2MASS) and the UKIRT
Infrared Deep Sky Survey (UKIDSS). NIR data are more suitable to study
OCs in the disc of our Milky Way galaxy than optical data, because NIR
observations can be used to limit the effects of differential
reddening. The limiting $J$-band magnitude of the 2MASS Point Source
Catalog \citep{Cutri03} is $J \approx 15.5$ mag, which is not
sufficiently faint for some clusters. Therefore, for faint clusters,
where available we added UKIDSS data to the 2MASS data at the faint
end. UKIDSS performed a northern sky survey of 7500 deg$^2$, reaching
up to 3 mag fainter than 2MASS. Specifically, the catalogue we used is
the UKIDSS Data Release 6 Galactic Plane Survey \citep{Lucas08}.

\subsection{The Cepheid catalogue}
\label{catalogues.sec}

The Cepheid catalogue contains the All Sky Automated Survey (ASAS)
Catalogue of Variable Stars, the General Catalogue of Variable Stars
(GCVS) and the catalogue published by \citet{Tammann03}. The latter
includes 324 Cepheids; it was derived from \citet{Berdnikov00} and a
systematic trend in colour excesses was removed by
\citet{Tammann03}. The \citet{Berdnikov00} catalogue is a homogeneous
database containing $BVI$ photometry in the Cape (Cousins) photometric
system for hundreds of Galactic Cepheids; the colour excesses,
$E(B-V)$, are from \citet{Fernie94} and \citet{Fernie95}. The
\citet{Tammann03} catalogue contains good reddening values and
distances, so this catalogue has been used to check the confidence of
our reddening values, which we derived independently from isochrone
fitting (see Sect. 3.2). The ASAS Catalogue of Variable Stars contains
more than one thousand Cepheids, some of which are newly found or
poorly studied. Although only $V$-band data are available for most
Cepheids in this catalogue, its period information is good and it
contains complete light-curve data, which is sufficient to find new OC
Cepheids. The GCVS contains approximately 600 Cepheids.

From these databases we obtained positions for a list of Cepheids to
match to nearby OCs. Cepheid distances can be estimated using the
Cepheid PLR from \citet{Sandage06} if the reddening, apparent
magnitude and period ($P$) are known, giving absolute magnitudes, $M_V
= (3.087 \pm 0.085) \log(P \mbox{ day}^{-1}) + (0.914 \pm 0.098)$
mag. We compared the Cepheid and OC distances to constrain our sample
of possible OC Cepheids. To limit the effects of reddening, we used
$JHK$-band data to obtain the mean magnitudes for a subset of the
Cepheids. $JHK$-band mean magnitudes in the South African Astronomical
Observatory (SAAO) system are available for 229 Cepheids \citep{van
  Leeuwen07}. We transformed their photometry to that of the 2MASS
$JHK_{\rm s}$ photometric system using the recommended transformation
equations.\footnote{http://www.ipac.caltech.edu/2mass/releases/allsky/doc/
  sec6\_4b.html} The differences between the OC NIR colour excesses
and the Cepheids' optical colour excesses are small and the prevailing
extinction law \citep{Rieke85} is well-understood; $E(J-H) = 0.33
E(B-V)$. These OC NIR colour excesses can be adopted to help estimate
the Cepheid distances.

\subsection{Position selection}

A position cross-match was done between the OC centres and Cepheid
positions. Approximately 2100 OCs and more than 1000 Cepheids were
used. We adopted a maximum (projected) separation of 60 arcmin to
select OC--Cepheid combinations. This resulted in 600 possible
associations. The probability of finding genuine OC Cepheids at
distances in excess of one degree from a given cluster centre is very
small or even negligible. We first concentrated on the roughly one
hundred Cepheids which are located within 20 arcmin from the nearest
cluster centres and performed proper-motion and distance selection for
each object individually. For the Cepheids which were located at
20--60 arcmin from their associated cluster centres, we first compared
the magnitude of the cluster stars and the Cepheids in the context of
the cluster CMDs. We excluded those Cepheids which were fainter than
the equivalent cluster stars. Faint Cepheids are unlikely OC Cepheid
candidates, because in a given stellar population Cepheids are
brighter than the associated main-sequence stars.

\subsection{Distance selection}

Combined with position selection, distance selection is a very
effective approach to select OC Cepheids. If the distance and position
of a star are known, its 3D position can be determined. To perform
distance selection, we need to calculate the distances to our sample
Cepheids and OCs independently. For the Cepheids, we used the best
observational PLR to calculate their absolute $V$-band magnitudes
\citep{Sandage06}. The $V$-band mean apparent magnitude is available
for all of our Cepheids, so if we know the reddening, we can derive
the DM for the Cepheid of interest. \citet{Fernie94} provided $E(B-V)$
for some Cepheids. For the others, we used the $E(J-H)$ colour excess
based on our isochrone fits and converted it to $E(B-V)$. Although
this step may introduce a significant uncertainty (because the visible
reddening affecting the Cepheids and OCs may not be the same), this
will not prevent us from performing a robust distance selection and
thus finding possible OC Cepheids. This is so, because across the
small OC fields, the differential change in the associated colour
excess is much smaller than our distance selection criterion (see
below). OC DMs were obtained from isochrone fitting, independent of
the distances to the Cepheids. The typical apparent diameter of many
OCs is of order 10 arc\-min. However, some very bright OC Cepheids can
be located at 30 arcmin or more from the cluster centres.

We thus obtained a list of probable OC Cepheids, ordered by their
projected separation to the corresponding cluster centres. For
separations greater than 20 arcmin, Cepheids that were too faint were
removed (see Section 2.3). If they are fainter than the cluster's
main-sequence stars, these objects must be background
Cepheids. Inspection of finding charts (using the {\sc Aladin} viewing
tool) and 2MASS magnitudes were also used to exclude these bright
OC/faint Cepheid associations.

Because of the relatively bright limiting magnitude and the limited
photometric precision of 2MASS, the CMDs of some 100 OCs do not
exhibit tight main sequences, which is particularly an issue for
clusters with small numbers of members. We excluded these clusters
from our analysis, since the error in the DM $> 1$ mag. We imposed
$|{\rm DM}_{\rm clus}-{\rm DM}_{\rm Cep}| \le 1.0$ mag to compare the
DMs of our OC--Cepheid combinations and select possible OC
Cepheids. The distances to Galactic OCs are mostly 500--3000 pc, or
$(m-M)_0 = 8.5$--12.5 mag, so that if the difference in DMs between a
cluster and an apparently associated Cepheid is greater than 1.0 mag,
the probability that these objects are physically associated with one
another is very small indeed. Upon completion of this step, our sample
contained 45 possible OC--Cepheid associations.

\begin{figure*}
\begin{minipage}{120mm}
\includegraphics[width=120mm]{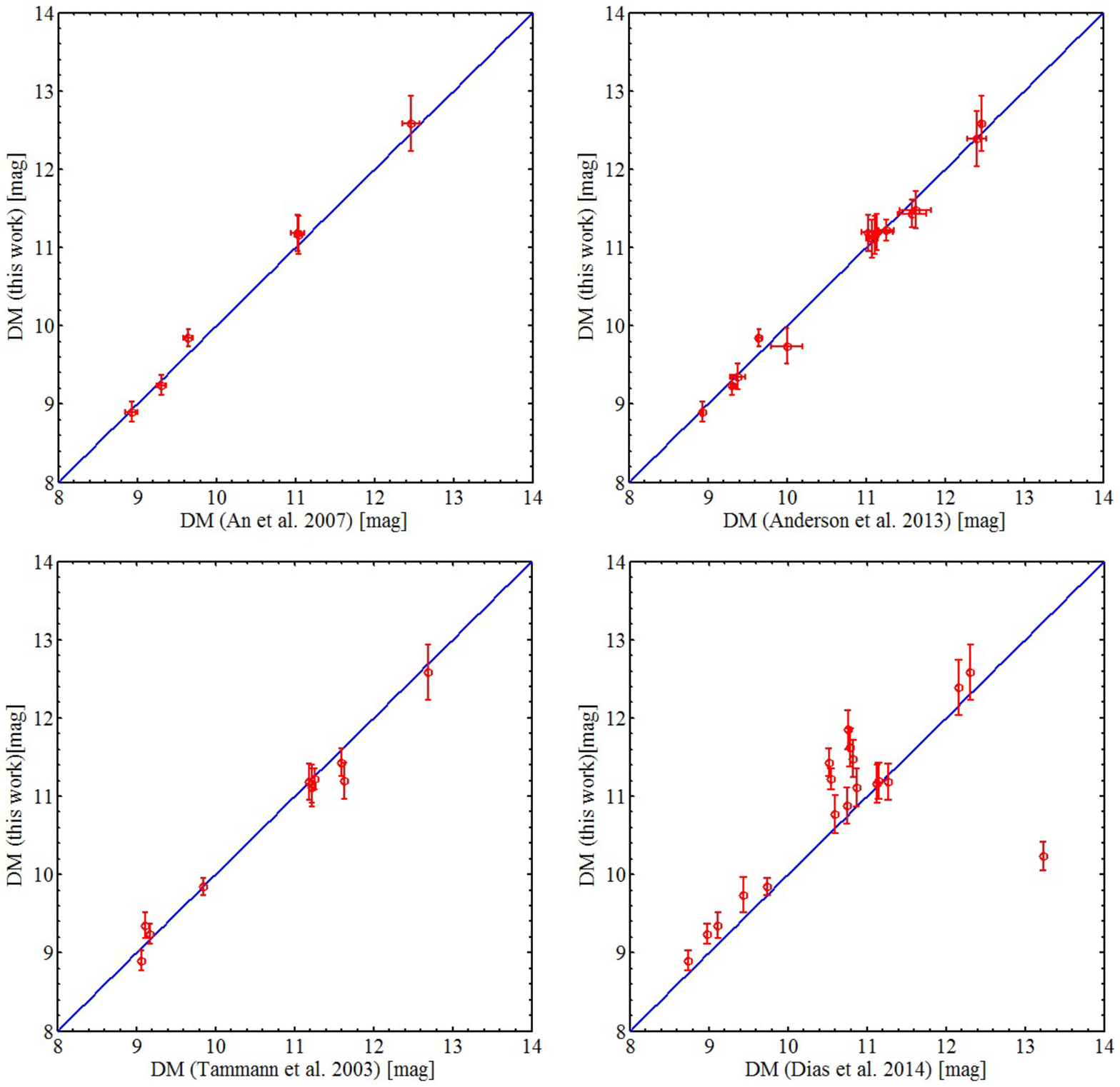}
\caption{Comparison of our newly derived distances with those of
  \citet{An07}, \citet{Anderson13}, \citet{Tammann03} and
  \citet{Dias14}.}
  \label{f0.fig}
  \end{minipage}
\end{figure*}

 To confirm the reliability of our distance
  estimates based on isochrone fitting, in Table 1 and
  Fig. \ref{f0.fig} we compare our newly derived distances with those
  of \citet{An07}, \citet{Anderson13}, \citet{Tammann03} and
  \citet{Dias14}. \citet{An07} used a multi-passband $BVI_{\rm
    C}JHK_{\rm s}$ combination to derive their distance estimates.
  For six of their OCs in common with our sample, our distance
  estimates fall at (for NGC 6087) or within the mutual 1$\sigma$
  uncertainties.\footnote{We do not consider our distance to Lyng{\aa}
    6 as very reliable because of the small number of points in the
    2MASS CMD this value is based on; \citet{An07} do not provide a
    $JHK_{\rm s}$-based distance for the same reason.}
  \citet{Anderson13} provide an up-to-date summary of literature-based
  distance estimates. A comparison of our distances with their best
  estimates, which are based on careful analysis of the uncertainties
  affecting the various literature values, shows excellent mutual
  consistency for each OC in common. Similarly, a comparison of our
  results with the homogeneous data set of \cite{Tammann03}
  (cf. Section \ref{catalogues.sec}) shows a negligible mean
  difference of $-0.07 \pm 0.17$ mag (in the sense of the Tammann
  catalogue data minus our values), where the uncertainty represents
  the standard deviation of the data points. Finally, the
  \citet{Dias14} catalogue includes distances to $94.0\%$ of Galactic
  OCs. Although this data set therefore serves as a useful comparison
  benchmark, these latter authors do not provide uncertainties and
  some of their isochrone fits appear relatively poor. Figure
  \ref{f0.fig} suggests a small but systematic offset of $0.35 \pm
  0.34$ mag between the Dias et al. (2014) distances and our results
  and, by extension, with respect to the other comparison samples. The
  latter are all consistent with our results within the 1$\sigma$
  uncertainties (with the exception of ASCC 61).

\begin{table*}
 \begin{minipage}{120mm}
\caption{Comparison of OC absolute distance moduli}
\begin{tabular}{@{}lrrrrrccc@{}}
  \hline
 Open cluster & \multicolumn{1}{c}{This paper} & \multicolumn{1}{c}{An et al. (2007)} & \multicolumn{1}{c}{Anderson} & \multicolumn{1}{c}{Tammann} & \multicolumn{1}{c}{Dias et} \\
 &  &  & \multicolumn{1}{c}{et al. (2013)} & \multicolumn{1}{c}{et al. (2003)} & \multicolumn{1}{c}{al. (2014)} \\
  \hline
Dolidze 34    & $11.85\pm0.25$ &                &                &          & 10.76\\
Dolidze 52    & $10.77\pm0.24$ &                &                &          & 10.60\\
NGC 6683      & $11.62\pm0.24$ &                &                &          & 10.79\\
ASCC 61       & $11.19\pm0.23$ &                & $11.14\pm0.20$ &  11.63   & 11.15\\
Feinstein 1   & $10.88\pm0.23$ &                &                &          & 10.75\\
NGC 7790      & $12.58\pm0.35$ & $12.46\pm0.11$ & $12.46\pm0.01$ &  12.69   & 12.31\\
NGC 6087      & $9.84\pm0.11$  & $ 9.65\pm0.06$ & $9.65\pm0.03$  &   9.85   &  9.74\\
IC 4725       & $8.90\pm0.13$  & $ 8.93\pm0.08$ & $8.93\pm0.02$  &   9.07   &  8.74\\
vdBergh 1     & $11.11\pm0.24$ &                & $11.08\pm0.07$ &  11.22   & 10.87\\
NGC 129       & $11.16\pm0.24$ & $11.04\pm0.05$ & $11.11\pm0.02$ &  11.22   & 11.13\\
Collinder 394 & $9.35\pm0.16$  &                & $9.38\pm0.10$  &   9.11   &  9.12\\
Turner 2      & $11.22\pm0.13$ &                & $11.26\pm0.10$ &  11.26   & 10.55\\
Trumpler 35   & $11.43\pm0.18$ &                & $11.58\pm0.18$ &  11.60   & 10.52\\
Ruprecht 175  & $10.23\pm0.18$ &                &                &          & 13.23\\
NGC 5662      & $9.24\pm0.13$  & $ 9.31\pm0.06$ & $9.31\pm0.02$  &   9.17   &  8.98\\
ASCC 69       & $9.74\pm0.23$  &                & $10.0\pm0.20$  &          &  9.44\\
Collinder 220 & $11.48\pm0.24$ &                & $11.63\pm0.20$ &          & 10.83\\
NGC 6067      & $11.18\pm0.23$ & $11.03\pm0.08$ & $11.03\pm0.01$ &  11.19   & 11.27\\
Berkeley 58   & $12.39\pm0.35$ &                & $12.40\pm0.12$ &          & 12.16\\
\hline
 \end{tabular}
\end{minipage}
\end{table*}

\subsection{Proper-motion selection}

Proper-motion selection is a powerful tool to separate OC members from
foreground or background stars. We compared the average cluster proper
motions with those pertaining to the apparently associated Cepheids to
further constrain our OC Cepheid sample. The proper-motion data we
used were obtained from the PPMXL Catalogue \citep{Roeser10}, which is
a full-sky survey down to $V = 20$ mag, containing some 900 million
objects. 2MASS contains about 400 million objects; the PPMXL Catalogue
provides proper-motion data for all 2MASS objects. The resulting,
typical individual mean errors of the proper motions range from 4 mas
yr$^{-1}$ to more than 10 mas yr$^{-1}$. The \hbox{PPMXL} Catalogue
provides an effective way to study cluster membership because of the
large database that reaches sufficiently faint magnitudes.

In order to reduce contamination by non-members, we first plotted the
CMDs of our OCs and selected stars along their main sequences for our
proper-motion selection (see Figure \ref{f2.fig}; we excluded stars characterized
by $J-H < 0.2$ mag and $J-H >0.8$ mag). Next, we plotted the
proper-motion distribution of the OC of interest; the locus dominated
by the cluster members is usually very obvious. We used these stars to
obtain an initial estimate of the average proper motion for the
cluster and its $1 \sigma$ standard deviation, so as to select genuine
OC members. We then reapplied our proper-motion selection to all stars
in the cluster area to derive $\mu_{\alpha,{\rm cl}}$ (in right
ascension) and $\mu_{\delta,{\rm cl}}$ (in declination) more robustly
for the cluster as a whole. We compared the Cepheid's proper motion,
$(\mu_{\alpha,{\rm Cep}},\mu_{\delta,{\rm Cep}})$, with
$(\mu_{\alpha,{\rm cl}}, \mu_{\delta,{\rm cl}})$ to assess the
Cepheid's membership probability. We excluded Cepheids that are
located beyond $2 \sigma$ from the cluster's average proper
motion. Figure \ref{f1.fig} shows an example of our proper-motion
selection (see the Appendix for the relevant figures for all other
OC--Cepheid combinations discussed in this paper). The black dots are
the stars selected after the first application of our selection
criteria, while the blue dots are the stars located within the $1
\sigma$ range after the second selection step. The red triangles
represent the Cepheids' proper motions. We compared the loci of the
red triangles to the range of blue dots to select the most likely
OC Cepheids. Cepheids whose proper motions result in a locus between
$1 \sigma$ and $2 \sigma$ were considered possible OC Cepheids. Upon
application of this selection procedure, we were left with 37
potential OC Cepheids.

\begin{figure}
\centering
\includegraphics[width=85mm]{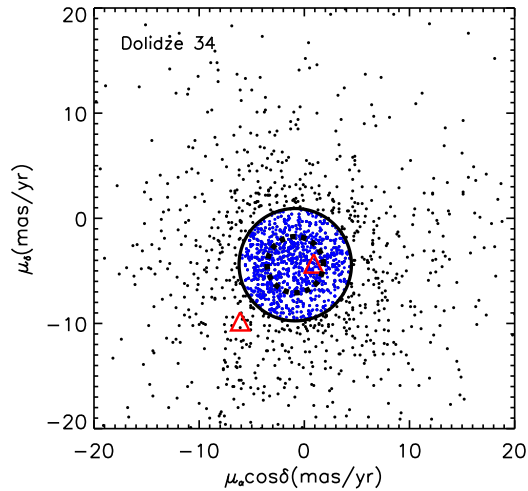}
\caption{Example of our proper-motion selection. The black dots are
  the stars selected after the first application of our selection
  criteria, while the blue dots are the stars located within the $1
  \sigma$ radius after the second selection step. The red triangles
  represent the Cepheids' proper motions.  The
    dotted and solid circles represent the 0.5$\sigma$ and 1$\sigma$
    radii, respectively. We compared the loci of the red triangles to
  the range of blue asterisks to select the most likely OC
  Cepheids. One of the Cepheids in this example is located inside the
  $1 \sigma$ radius, whereas the other Cepheid's locus is found at
  approximately $1.4 \sigma$. The latter object is considered a
  possible OC Cepheid.}
  \label{f1.fig}
\end{figure}

\subsection{Instability-strip selection}

\begin{figure}
\centering
\includegraphics[width=75mm]{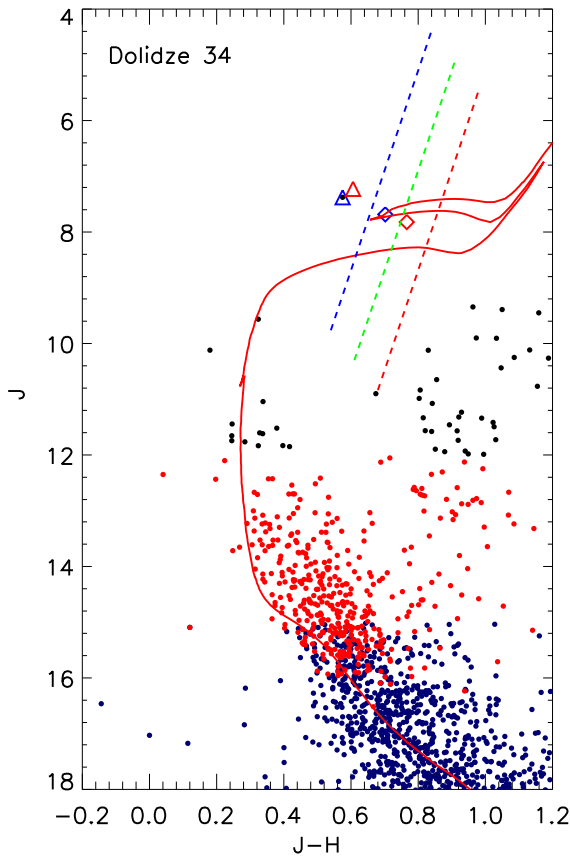}
\caption{Example of isochrone fitting to the CMD of the OC Dolidze 34
  (cf. Fig. \ref{f1.fig}). The black dots are stars brighter than
  $J_{\rm 2MASS} = 12$ mag. The red dots are from the UKIDSS database,
  showing stars fainter than $J_{\rm 2MASS} = 12$ mag, down to the
  limiting magnitude of the proper-motion catalogue. The 2MASS and
  UKIDSS cluster stars were selected based on proper-motion
  constraints. The blue dots are fainter stars located within 3 arcmin
  from the cluster centre, which we use to show a continuous main
  sequence. The solid red line is the best-fitting isochrone for solar
  metallicity. The green dashed line is the central ridge line of the
  Cepheid instability strip from \citet{Tammann03}; the red and blue
  dashed lines are the red and blue edges of the Cepheid instability
  strip, respectively. The blue diamond and triangle show the loci of
  two Cepheids in the 2MASS catalogue, while the red symbols represent
  their mean $(J-H)$ colour and $J$ magnitude,
  respectively.\label{f2.fig}}
\end{figure}

OC Cepheids must be located in the instability strip of their host
clusters. Figure \ref{f2.fig} shows the Cepheid instability strip's
centre, as well as the corresponding red and blue edges, using the
photometric conversion of \citet{Sandage08}. The centre position (the
`ridge line') is based on the observed Cepheid instability strip,
$\log T_{\rm eff} = -0.054 \log L_V + 3.922$ \citep{Sandage04}. The
blue and red edges have been drawn arbitrarily using an
effective-temperature width of $\Delta \log T_{\rm eff} = 0.06$ [K],
which is slightly wider than what is observed \citep[cf.][their
  fig. 20]{Sandage04}. We first converted the luminosity $L$ to
$M_{J}$, adopting $M_{J,\odot}=3.67$ mag and a bolometric correction,
$BC \sim 0$ mag; $BC \ll 0.1$ mag \citep[cf.][their table
  6]{Sandage99}. Next, we converted $T_{\rm eff}$ to $(J-H)$, using
Padova stellar spectra \citep{Bressan12} and adopting $\log T_{\rm
  eff} = -0.387(J-H) + 3.886$ [K]. Thus, the ridge line shown in our
CMDs is given by $M_{J} = -13.810(J-H) + 3.118$ mag, which we combined
with the apparent distance and an appropriate reddening correction for
each OC.

We compared the positions of the Cepheids and the instability strip
for each OC to assess their membership probabilities. We adopted the
blue and red edges as a proxy to the $1 \sigma$ deviations. Cepheids
found beyond $2 \sigma$ from the ridge line were excluded from further
consideration. Since the magnitudes and colours of Cepheids change
periodically, we used the mean magnitudes and colours. We added the
mean magnitudes of the Cepheids used in our example cluster
(cf. Fig. \ref{f1.fig}) as red diamonds to Fig. \ref{f2.fig}. All
Cepheids selected in our previous selection steps satisfy our
instability-strip selection criteria.

\subsection{Age selection}

\begin{figure}
\centering
\includegraphics[width=85mm]{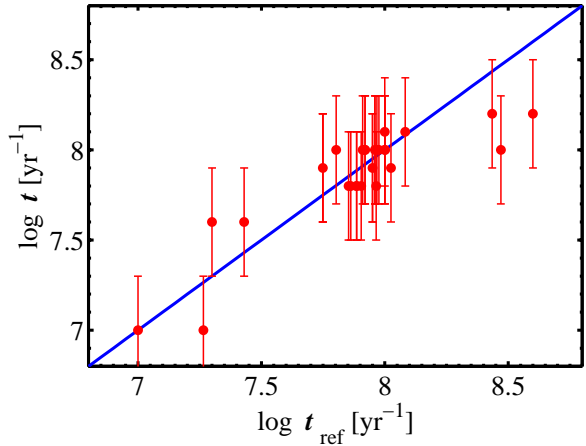}
\caption{Comparison of our derived ages with those taken from the
  WEBDA database.\label{f3new.fig}}
\end{figure}

The ages of cluster Cepheids should be comparable to those of their
host OCs. Although this is not necessarily a hard-and-fast rule,
matching Cepheid and OC ages provides at least some moderately useful
additional constraints. All Galactic OCs are young, with ages usually
between $10^7$ and $10^9$ yr. Cepheid ages can be calculated based on
theoretical pulsation modelling \citep{Bono05}. For fundamental-mode
Cepheids, $\log( t \mbox{ yr}^{-1}) = (8.31 \pm 0.08) - (0.67 \pm
0.01) \log (P \mbox{ day}^{-1})$, while for first-overtone Cepheids,
$\log( t \mbox{ yr}^{-1}) = (8.08 \pm 0.08) - (0.39 \pm 0.04) \log (P
\mbox{ day}^{-1})$. Fundamental-mode and first-overtone Cepheids have
ages around $\log( t \mbox{ yr}^{-1}) = 7.6$ at $\log (P \mbox{
  day}^{-1})=1.0$. OC ages are usually calculated based on isochrone
fitting. We used the ages listed for some OCs in the WEBDA database as
reference to check the reliability of our isochrone fits (see
Fig. \ref{f3new.fig}).

The age uncertainties associated with isochrone fitting can be of
order $\Delta \log( t \mbox{ yr}^{-1}) = 0.2$--0.3 for OC ages around
$10^8$ yr. We adopted as age-selection criterion an age difference
between the Cepheids and their postulated host OCs of $\Delta \log( t
\mbox{ yr}^{-1}) \le 0.3$; if $\Delta \log( t \mbox{ yr}^{-1}) > 0.3$,
a more careful check of our CMDs would be required. All 37 OC Cepheids
in our constrained sample satisfy this criterion.

\subsection{Radial-velocity and iron-abundance selection}

Additional selection criteria applied to our sample of potential OC
Cepheids include radial-velocity and iron-abundance selection. Radial
velocities and iron-abundance data are available for only a few OCs in
the catalogues. The DAML02 (version 3.4) radial-velocity data we used
for our OCs come mostly from \citet{Dias14} and \citet{Kharchenko13},
whereas the radial velocity data for the Cepheids are from
\citet{Storm11} and \citet{Fernie95}. Since more than half of the OCs'
radial velocities are calculated based on fewer than three stars, we
used $\sigma_{\rm RV}=10/ (\sqrt N)$ km s$^{-1}$ as our selection
criterion \citep{Anderson13}.

\citet{Anderson13} carried out a detailed analysis of every
OC--Cepheid combination available to them. Therefore, we use
their-iron abundance results as reference in the next section.

\section{Results}

\subsection{CMD analysis}

CMD analysis was done for each OC in our sample. We used 2MASS
$JHK_{\rm s}$ photometry for our colour--magnitude analysis, applying
an additional selection criterion in which we only considered stars
with photometric errors of less than 0.2 mag. For a number of our OCs,
UKIDSS data are also available. In these cases, 2MASS data were
adopted for stars brighter than $J = 12$ mag, and UKIDSS data for
fainter stars. We converted the UKIDSS $JHK$ photometry to 2MASS
$JHK_{\rm s}$ magnitudes using \citep{Hewett06}
\begin{eqnarray}
J_{\rm 2MASS} &=& J-0.01+0.073(J-H); \nonumber \\
H_{\rm 2MASS} &=& H-0.069(H-K); \\
K_{\rm 2MASS} &=& K+0.073(H-K). \nonumber
\end{eqnarray}

These conversion equations apply to main-sequence stars. For those OCs
whose CMDs revealed well-defined cluster sequences, we determined the
distances, reddening values and ages by fitting the cluster sequences
with Padova isochrones \citep{Bressan12} in the 2MASS $JHK_{\rm s}$
photometric system. If the cluster metallicity was not known, we
adopted solar-metallicity isochrones. This method is identical to that
adopted by \citet{Kronberger06}, who used isochrone fitting to find
new OCs in the 2MASS catalogue. We also used the $(J-H)$ vs $(H-K_{\rm
  s})$ colour--colour diagram to check the validity of our reddening
values (see Section 3.2). In Figure \ref{f2.fig} we show an example of
an isochrone-fitting result based on a $J$ vs $(J-H)$ CMD (see the
Appendix for the equivalent figures for our other OCs). From isochrone
fits to this particular CMD we determined the cluster parameters,
$\log(t \mbox{ yr}^{-1})=7.9 \pm 0.3, E(J-H)=0.36\pm0.02$ mag and an
apparent DM of $(m-M)_{J}=12.80 \pm 0.20$ mag. Using the relevant
extinction law \citep{Rieke85}, $A_J=2.64 E(J-H)$, the absolute DM
yields $(m-M)_{0,J}=11.85 \pm 0.25$ mag.

\subsection{Reddening check}

\begin{figure}
\centering
\includegraphics[width=85mm]{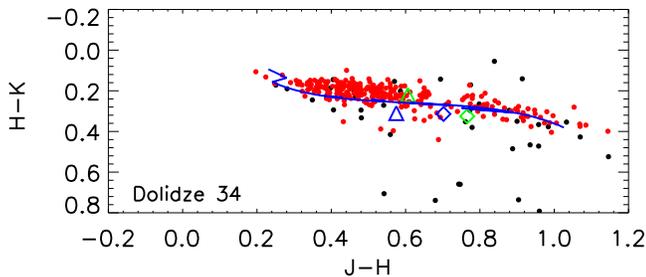}
\caption{Colour--colour diagram of Dolidze 34. The black and red dots
  are from 2MASS and UKIDSS, respectively, based on proper-motion
  selection. To exclude field stars we only include stars with $J<14$
  mag. The blue line is derived from the intrinsic colours adopted
  \citep{Turner11} and the NIR extinction law of \citet{Wang14},
  giving a colour excess of $E(J-H)=0.36$ mag. The blue symbols show
  the loci of the Cepheids in this cluster based on the 2MASS
  catalogue, and the green symbols show their average
  magnitudes. \label{f3.fig}}
\end{figure}

Correcting for reddening is very important to improve the accuracy of
our OC properties. If we know a cluster's reddening and age, we can
accurately derive its distance based on isochrone
fitting. Colour--colour diagrams were used to obtain the appropriate
colour excesses. For main-sequence cluster stars, we took the
intrinsic colours, $(J-H)_0$ and $(H-K_{\rm s})_0$, from
\citet{Turner11}. \citet{Wang14} determined a universal NIR extinction
law for the Galactic plane, $E(J-H)/E(J-K_{\rm s})=0.64,
E(J-H)/E(H-K_{\rm s})=1.78$. Combining the intrinsic colours with this
extinction law, we derived the $(J-H)$ vs $(H-K_{\rm s})$
relationship, shown as the blue line in the colour--colour diagram of
Fig. \ref{f3.fig}. We found that this relationship represents the
ridge line defined by our data points. As an example, Fig.
\ref{f3.fig} illustrates this procedure for Dolidze 34, which is
characterized by a colour excess, $E(J-H)=0.36 \pm 0.02$ mag. (See the
Appendix for the equivalent figures for our other sample OCs.)

\begin{figure}
\centering
\includegraphics[width=85mm]{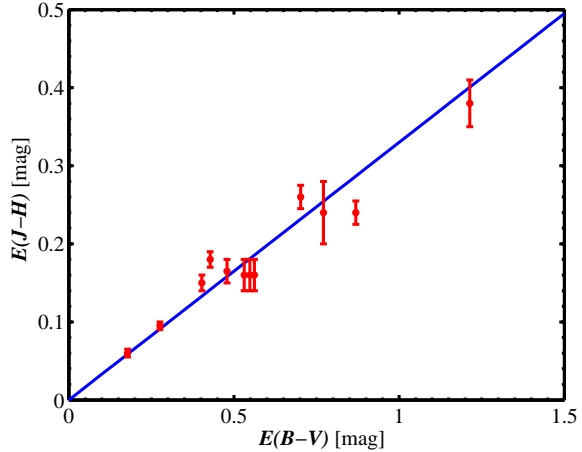}
\caption{Comparison of our newly determined NIR reddening values with
  their optical counterparts from \citet{Tammann03}. The solid line
  represents the relation, $E(J-H) = 0.33E(B-V)$.
  \label{f4.fig}}
\end{figure}

If we assume that most OCs are negligibly affected by differential
reddening effects because of their small sizes, the associated OC
Cepheids will -- to first order -- have the same reddening. The
\citet{Tammann03} Cepheid catalogue includes $E(B-V)$ reddening
corrections, while we obtained the equivalent NIR values for our OCs
based on isochrone fitting. The reddening corrections in
\citet{Tammann03} were derived from \citet{Fernie94} and
\citet{Fernie95}, which represents the largest body of $E(B-V)$ values
for Galactic Cepheids available to date. \citet{Tammann03} applied a
small systematic correction, $E(B-V)_{\rm corr} =
(0.951\pm0.012)E(B-V)_{\rm Fernie}$. We converted our $E(J-H)$
reddening estimates to $E(B-V)$ values using $E(J-H) = 0.33E(B-V)$
\citep{Rieke85}. Figure \ref{f4.fig} shows our CMD-based reddening
estimates vs Cepheid reddening values from \citet{Tammann03}. This
figure shows that our reddening estimates are robust.

\subsection{Possible open cluster Cepheids}

\begin{figure}
 \centering
\includegraphics[width=85mm]{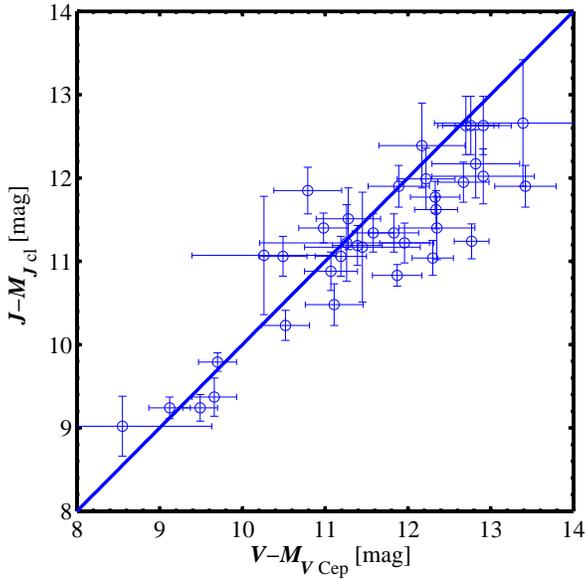}
\caption{Cepheid vs OC DMs. The latter were estimated based on
  isochrone fitting, while Cepheid distances were derived on the basis
  of the observational PLR. Both have been reddening-corrected. The
  solid line shows the one-to-one relation. All 37 Cepheids meeting
  our distance criterion are included. \label{f5.fig}}
\end{figure}

Application of our five selection criteria resulted in a sample of 37
possible OC Cepheids. Note that we have not included a number of
probable first-overtone Cepheids such as QZ Nor and V1726 Cyg
\citep{Tammann03}. Figure \ref{f5.fig} shows a comparison of Cepheid
and OC DMs. The solid line is the one-to-one relation. The error bars
associated with the distance estimates based on our isochrone fits
contain the uncertainties in both the DM and the $J$-band extinction,
$A_J$. The error bars associated with the Cepheid distances include
the uncertainties in both the PLR and the $V$-band extinction. The
latter is often too large to lead to a reliable PLR.

\begin{figure}
\includegraphics[width=85mm]{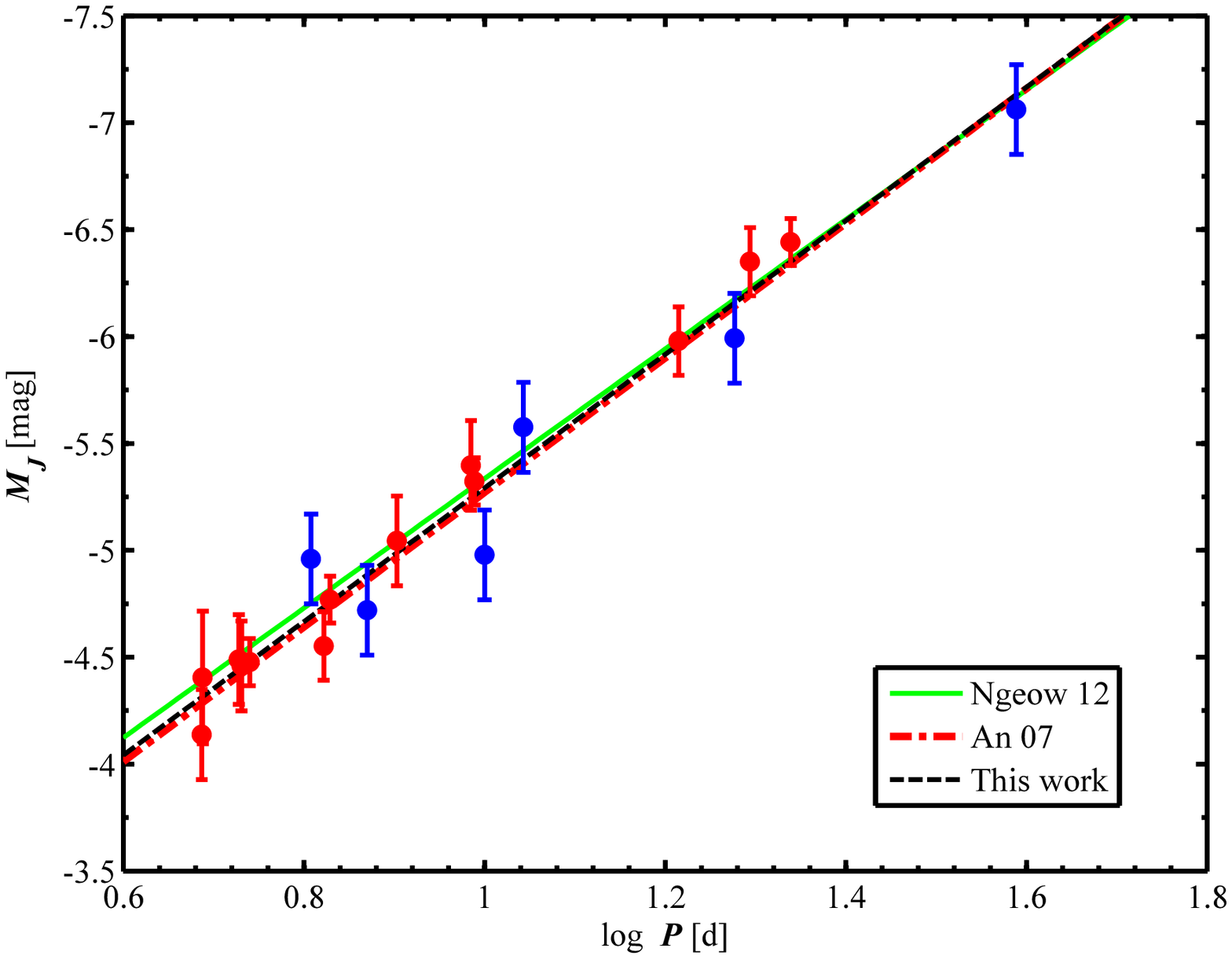}
\caption{$J$-band PLR for our 19 OC Cepheids. The red dots are the
  confirmed OC Cepheids, while the blue dots are the high-possibility
  OC Cepheid candidates. The green solid and red dash-dotted lines are
  the $J$-band PLR for Galactic Cepheids from \citet{Ngeow12} and
  \citet{An07}, respectively; the black dashed line is our linear fit
  result.\label{f6.fig}}
\end{figure}

To reduce the effects of differential reddening between the Cepheids
and their postulated host OCs, and to avoid introducing errors through
application of a reddening conversion from NIR to visible wavelengths,
we opted to use the NIR mean magnitudes. $JHK$-band mean magnitudes in
the SAAO system are available for 229 Cepheids \citep{van
  Leeuwen07}. Among our sample of possible OC Cepheids, these mean
magnitudes are available for only 22 of the 37 Cepheid candidates;
three of these 22 Cepheids were excluded because they did not satisfy
our NIR selection criteria (see below). The remaining 19 OC Cepheids
are included in Table 1 and their corresponding CMDs, colour--colour
and proper-motion diagrams are provided in the Appendix. The ages,
reddening values and DMs for these OCs based on isochrone fitting are
also included in Table 1. Thirteen of these 19 Cepheids are confirmed
OC Cepheids and the other six are very likely OC Cepheids, because
their distances and proper motions span the same ranges as those of
their host OCs. These latter Cepheids are TY Sct and CN Sct in Dolidze
34, XX Sgr in Dolidze 52, CK Sct in NGC 6683, VY Car in ASCC 61 and U
Car in Feinstein 1. Figure \ref{f6.fig} shows the PLR based on this
sample of OC Cepheids. The distance to each Cepheid was estimated
based on isochrone fitting. The red dots are the confirmed OC
Cepheids, while the blue dots are possible OC Cepheids. The green line
is the $J$-band PLR for Milky Way Cepheids, $M_J= (-3.030 \pm 0.022)
\log(P \mbox{ day}^{-1})-(2.306 \pm 0.020)$ mag \citep{Ngeow12},
 whereas the green line is the $J$-band PLR from
  \citet{An07}, $M_J= (-5.271 \pm 0.076) \log(P \mbox{ day}^{-1}
  -1)-(3.148 \pm 0.053)$ mag. Our new Cepheids conform with both of
these PLRs; our newly derived PLR is given by $M_J= (-3.12 \pm 0.29)
\log (P \mbox{ day}^{-1})-(2.17 \pm 0.29)$, with a correlation
coefficient, $R^2=0.9681$, and an r.m.s. of 0.15 mag.

\subsection{OC Cepheid candidates without NIR observations}

For completeness, Table 2 includes an overview of the selection
matches for the 15 OC Cepheid candidates for which we do not have
access to NIR data. We will now briefly discuss the extent to which
these objects are genuine OC Cepheids.

\subsubsection{Confirmed Cepheids}

\paragraph{Berkeley 58 and CG Cas}
\citet{Turner08} confirmed CG Cas as a member of Berkeley 58, although
their proper motions disagree to some extent: the proper motion of CG
Cas is found within the 1.5$\sigma$ range of genuine Berkeley 58
members. Its DM and reddening are $(m-M)_{0,V}=12.40 \pm 0.12$ mag and
$E(B-V)=0.70 \pm0.03$ mag, respectively, which is within 1--2$\sigma$
of our NIR determinations, i.e., $(m-M)_{0,J}=12.39 \pm 0.35$ mag and
$E(J-H)=0.23\pm0.02$ mag, respectively. The CG Cas--Berkeley 58
combination meets all four other selection criteria.

\paragraph{NGC 7790 and CEa Cas, CEb Cas}
NGC 7790 was studied in detail by \citet{An07}. They derived
$(m-M)_{0,V}=12.46 \pm 0.11$ mag and $E(B-V)=0.48 \pm 0.02$ mag,
compared with our determinations of $(m-M)_{0,J}= 12.58\pm0.31$ mag
and $E(J-H)=0.14\pm0.02$ mag, respectively. Like CF Cas, CEa Cas and
CEb Cas satisfy all five selection criteria for NGC 7790. CF Cas and
CEab Cas were shown to be NGC 7790 members by \citet{Majaess13}.

\paragraph{Kharchenko 3 and ASAS J182714$-$1507.1}
\citet{Anderson13} first suggested that ASAS J182714$-$1507.1 may be a
genuine member of the OC Kharchenko 3. These authors showed that the
ages, distances and proper motions of both Kharchenko 3 and ASAS
J182714$-$1507.1 are mutually consistent. \citet{Dias14} lists
$(m-M)_{0,V}=11.64$ mag and $E(B-V)=0.72$ mag, compared with our
determinations of $(m-M)_{0,J}= 11.51\pm0.37$ mag and
$E(J-H)=0.26\pm0.025$ mag, respectively. ASAS J182714$-$1507.1
satisfies all five criteria needed to confirm it as a member of
Kharchenko 3.

\subsubsection{New Cepheids}

\paragraph{NGC 6991 and V0520 Cyg}
V0520 Cyg is located at approximately 7 arcmin from the centre of NGC
6991. In the cluster's CMD, V0520 Cyg is found near the instability
strip's ridge line. We derive $E(J-H)=0.24 \pm 0.02$ mag, $\log(t
\mbox{ yr}^{-1}) = 8.2 \pm 0.3$ and $(m-M)_{0,J}=11.17 \pm 0.35$
mag. DAML02 lists $\log(t \mbox{ yr}^{-1}) = 9.11$ for NGC 6991 and an
apparent diameter of 24 arcmin. This close pair was overlooked for
many years because of errors affecting the OC
catalogue. \citet{Anderson13} excluded this OC--Cepheid pair from
their compilation because of the relatively large differences in ages,
proper motions and radial velocities contained in the old version of
the DAML02 database. However, the latest release of the database
includes improved proper-motion and radial-velocity measurements,
supporting our identification of NGC 6991 and V0520 Cyg as a genuine
OC--Cepheid pair.

The proper motion of NGC 6991 is $(\mu_{\alpha,{\rm
    cl}},\mu_{\delta,{\rm cl}}) = (-0.46 \pm 3.11, -1.54 \pm -3.23)$
mas yr$^{-1}$, which is in good agreement with $(\mu_{\alpha,{\rm
    ref}},\mu_{\delta,{\rm ref}}) = (-1.5, 1.94)$ mas yr$^{-1}$
\citep{Dias14}. The proper motion of V0520 Cyg, $(\mu_{\alpha,{\rm
    Cep}},\mu_{\delta,{\rm Cep}}) = (-1.6, -3.2)$ mas yr$^{-1}$, is in
excellent agreement with that of NGC 6991. The radial velocity of NGC
6991 is $v_{\rm r}=-21.77\pm5.77$ km s$^{-1}$ \citep{Dias14} -- based
on three stars -- which is comparable with that of V0520 Cyg, $v_{\rm
  r}=-23.0$ km s$^{-1}$. The apparent discrepancy in age between the
OC and the Cepheid is most likely owing to the difficulty in
performing robust isochrone fitting in the crowded 2MASS field.

\paragraph{Juchert 18 and CS Mon}
CS Mon is located at about 14 arcmin from the centre of Juchert 18. We
derive $E(J-H)=0.17 \pm 0.015$ mag, $\log(t \mbox{ yr}^{-1}) = 8.1 \pm
0.3$ and $(m-M)_{0,J}=12.35 \pm 0.24$ mag. Juchert 18 and CS Mon
perfectly satisfy our five selection criteria. The
\citet{Kharchenko13} isochrone-fitting result leads to a very old
cluster, $\log(t \mbox{ yr}^{-1}) = 9.02 \pm 0.04$. On the other hand,
when we use UKIDSS data instead of the shallower 2MASS data for the
faint stellar population, the CMD is better represented by that of a
young OC.

\subsubsection{Uncertain cluster Cepheids}

\paragraph{Trumpler 9 and ASAS J075503$-$2614.3}
\citet{Anderson13} suggested that the status of ASAS J075503$-$2614.3
as a member of Trumpler 9 is inconclusive, because their proper
motions and ages are discrepant. However, we find that the DMs and
ages both meet our selection criteria and that the proper motions
agree to within 1.6$\sigma$. It would be crucial to obtain the radial
velocities of both Trumpler 9 and ASAS J075503$-$2614.3 (although the
latter is rather faint) to ascertain the Cepheid's OC membership.

\paragraph{Ruprecht 65 and AP Vel}
Ruprecht 65 and AP Vel satisfy our proper-motion and age criteria. We
estimate $E(J-H)=0.20 \pm 0.03$ mag and $(m-M)_{0,J}=11.07 \pm 0.38$
mag, with an OC--Cepheid DM difference of 0.8 mag, which however
renders this a questionable OC--Cepheid pair.

\paragraph{Czernik 8 and UY Per}
\citet{Turner10} suggested that UY Per might be a member of Czernik
8. However, \citet{Anderson13} questioned this conclusion, since UY
Per's proper motion is much faster than that of the cluster. We
estimate a Cepheid proper motion of $(\mu_{\alpha,{\rm
    Cep}},\mu_{\delta,{\rm Cep}}) = (-6.1, 12.9)$ mas yr$^{-1}$, which
is 3.4$\sigma$ away from the mean proper motion of Czernik 8,
$(\mu_{\alpha,{\rm cl}}, \mu_{\delta,{\rm cl}}) = (-1.28 \pm 2.94,
-0.39 \pm 3.01)$ mas yr$^{-1}$. The DMs and ages satisfy our selection
criteria; the radial velocity of Czernik 8 is not available. We
conclude that UY Per's membership status of Czernik 8 is inconclusive.

\subsubsection{Rejected cluster Cepheids}

\paragraph{Ruprecht 79 and CS Vel}
The probability that CS Vel may be a member of Ruprecht 79 is less
than one per cent \citep{Anderson13}. Our results also show that the
ages, proper motions and radial velocities of Ruprecht 79 and CS Vel
do not agree very well. The proper motion of CS Vel is
$(\mu_{\alpha,{\rm Cep}},\mu_{\delta,{\rm Cep}}) = (-1.4, -8.6)$ mas
yr$^{-1}$, which is 2.0$\sigma$ away from the mean proper motion of
Ruprecht 79, $(\mu_{\alpha,{\rm cl}}, \mu_{\delta,{\rm cl}}) = (-5.66
\pm 4.67, 4.04 \pm 4.93)$ mas yr$^{-1}$. The radial velocity of CS Vel
is $v_{\rm r}=26.95$ km s$^{-1}$ \citep{Anderson13}, and that of
Ruprecht 79 is $v_{\rm r}=21.4\pm3.6$ km s$^{-1}$
\citep{Rastorguev99}. Finally, the age of Ruprecht 79 is $\log(t
\mbox{ yr}^{-1}) = 7.0 \pm 0.3$, which places it within the
2--3$\sigma$ range of that of CS Vel, $\log(t \mbox{ yr}^{-1}) = 7.8
\pm 0.1$.

\paragraph{Ruprecht 119 and ASAS J162811$-$5111.9}
Ruprecht 119 and ASAS J162811$-$5111.9 satisfy both our proper-motion
and age criteria. We estimate $E(J-H)=0.18 \pm 0.03$ mag and
$(m-M)_{0,J}=10.52 \pm 0.28$ mag. The OC--Cepheid DM difference is 0.7
mag, which casts doubt on the assertion that this may be a genuine
OC--Cepheid pair. The radial velocity of ASAS J162811$-$5111.9 is
4$\sigma$ from the average radial velocity of Ruprecht 119. ASAS
J162811$-$5111.9 is thus an unlikely member of Ruprecht 119.

\paragraph{Sher 1 and FN Car}
Sher 1 and FN Car satisfy the proper-motion and age criteria. We
estimate $E(J-H)=0.13 \pm 0.04$ mag and $(m-M)_{0,J}=12.66 \pm 0.51$
mag, but a DM difference of 0.7 mag. FN Car is located at about 7
arcmin from the centre of Sher 1; the apparent diameter of Sher 1 is 1
arcmin \citep{Dias14}. These results imply that FN Car is an unlikely
member of Sher 1.

\paragraph{ASCC 64 and HK Car}
Sher 1 and FN Car satisfy our proper-motion and age criteria. However,
the difference in DM is about 1.5 mag, which invalidates this
association as a genuine OC--Cepheid pair.

\paragraph{NGC 5999 and ASAS J155149$-$5621.8}
NGC 5999 and ASAS J155149$-$5621.8 satisfy the proper-motion
criterion. Based on a detailed analysis, we derive a difference in DM
of approximately 1.4 mag and and age difference, $\Delta \log(t \mbox{
  yr}^{-1}) = 0.4$, i.e., too large to conclusively identify this
association as a genuine OC--Cepheid pair.

\begin{table*}
 \begin{minipage}{120mm}
\caption{Isochrone-fitting results for our OC Cepheid candidates}
\begin{tabular}{@{}lccrrrccc@{}}
  \hline
  Cluster & $E(J-H)$ & $\log( t )$ & \multicolumn{1}{c}{$\mu$} & \multicolumn{1}{c}{$\mu_{0}$} & Cepheid  & $\langle J \rangle$ & $M_J$ & $\log( P )$\\
          & (mag)    & [yr]        & (mag)  & \multicolumn{1}{c}{(mag)}     &          & (mag) & (mag) & [day] \\
  \hline
Dolidze 34    & $0.36\pm0.02$   & 7.9 & 12.8 & $11.85\pm0.25$ & TY Sct & 7.225 & $-5.575$ & 1.043\\
Dolidze 34    & $0.36\pm0.02$   & 7.9 & 12.8 & $11.85\pm0.25$ & CN Sct & 7.821 & $-4.979$ & 1.000\\
Dolidze 52    & $0.24\pm0.015$  & 7.6 & 11.4 & $10.77\pm0.24$ & XX Sgr & 6.441 & $-4.959$ & 0.808\\
NGC 6683      & $0.18\pm0.015$  & 8.2 & 12.1 & $11.62\pm0.24$ & CK Sct & 7.380 & $-4.720$ & 0.870\\
ASCC 61       & $0.08\pm0.01$   & 8.0 & 11.4 & $11.19\pm0.23$ & VY Car & 5.408 & $-5.992$ & 1.277\\
Feinstein 1   & $0.12\pm0.01$   & 7.0 & 11.2 & $10.88\pm0.23$ & U Car  & 4.138 & $-7.062$ & 1.589\\
NGC 7790      & $0.14\pm0.02$   & 7.9 & 13.0 & $12.58\pm0.35$ & CF Cas & 8.595 & $-4.405$ & 0.688\\
NGC 6087      & $0.06\pm0.005$  & 8.0 & 10.0 & $9.84\pm0.11$  & S Nor  & 4.678 & $-5.322$ & 0.989\\
IC 4725       & $0.18\pm0.010$  & 7.8 &  9.3 & $8.90\pm0.13$  & U Sgr  & 4.531 & $-4.769$ & 0.829\\
vdBergh 1     & $0.26\pm0.015$  & 7.9 & 11.8 & $11.11\pm0.24$ & CV Mon & 7.341 & $-4.459$ & 0.731\\
NGC 129       & $0.165\pm0.015$ & 7.8 & 11.6 & $11.16\pm0.24$ & DL Cas & 6.556 & $-5.044$ & 0.903\\
Collinder 394 & $0.095\pm0.005$ & 8.0 &  9.6 & $9.35\pm0.16$  & BB Sgr & 5.048 & $-4.552$ & 0.822\\
Turner 2      & $0.18\pm0.01$   & 8.0 & 11.7 & $11.22\pm0.13$ & WZ Sgr & 5.258 & $-6.442$ & 1.339\\
Trumpler 35   & $0.33\pm0.01$   & 7.8 & 12.3 & $11.43\pm0.18$ & RU Sct & 5.950 & $-6.350$ & 1.294\\
Ruprecht 175  & $0.065\pm0.01$  & 7.8 & 10.4 & $10.23\pm0.18$ & X Cyg  & 4.421 & $-5.979$ & 1.215\\
NGC 5662      & $0.10\pm0.01$   & 8.0 &  9.5 & $9.24\pm0.13$  & V Cen  & 5.023 & $-4.477$ & 0.740\\
ASCC 61       & $0.08\pm0.01$   & 8.0 & 11.4 & $11.19\pm0.23$ & SX Car & 7.262 & $-4.138$ & 0.687\\
ASCC 69       & $0.06\pm0.01$   & 8.0 &  9.9 & $9.74\pm0.23$  & S Mus  & 4.503 & $-5.397$ & 0.985\\
Collinder 220 & $0.12\pm0.015$  & 8.1 & 11.8 & $11.48\pm0.24$ & UW Car & 7.311 & $-4.489$ & 0.728\\
\hline
 \end{tabular}
\end{minipage}
\end{table*}

\begin{table*}
 \centering
 \begin{minipage}{85mm}
\caption{Constraints for OC Cepheid candidates without NIR
  data. `$1\sigma$': within $1\sigma$; other values refer to the
  specific loci.}
\begin{tabular}{@{}llcccc@{}}
  \hline
  Cluster    &       Cepheid       &     DM    &     PM      &    RV     &    Age    \\
  \hline
Trumpler 9   & ASAS J075503$-$2614.3 & $1\sigma$ & $1.7\sigma$ &             & $1\sigma$ \\
Lynga 6      & TW NOR              &             & $1\sigma$   &             & $1\sigma$ \\
NGC 6991     & V0520 Cyg           & $1\sigma$   & $1\sigma$   & $1\sigma$   & $1\sigma$ \\
Sher 1       & FN Car              & $1\sigma$   & $1\sigma$   &             & $1\sigma$ \\
Ruprecht 65  & AP Vel              & $1\sigma$   & $1\sigma$   &             & $1\sigma$ \\
Juchert 18   & CS Mon              & $1\sigma$   & $1\sigma$   &             & $1\sigma$ \\
Ruprecht 119 & ASAS J162811$-$5111.9 & $1\sigma$ & $1\sigma$   & $4.1\sigma$ & $1\sigma$ \\
NGC 5999     & ASAS J155149$-$5621.8 & $1.4\sigma$ & $1\sigma$ &             & $1.2\sigma$ \\
Czernik 8    & UY Per              & $1\sigma$   & $3.4\sigma$ &             & $1\sigma$ \\
Berkeley 58  & CG Cas              & $1\sigma$   & $1.4\sigma$ &             & $1\sigma$ \\
NGC 7790     & CEA Cas             & $1\sigma$   & $1\sigma$   & $1\sigma$   & $1\sigma$ \\
NGC 7790     & CEB Cas             & $1\sigma$   & $1\sigma$   & $1\sigma$   & $1\sigma$ \\
Ruprecht 79  & CS Vel              & $1\sigma$   & $2.0\sigma$ & $1.5\sigma$ & $2.6\sigma$ \\
Kharchenko 3 & ASAS J182714$-$1507.1 & $1\sigma$ & $1\sigma$   &             & $1\sigma$ \\
ASCC 64      & HK Car              & $1.5\sigma$ & $1\sigma$   &             & $1\sigma$ \\
\hline
 \end{tabular}
\end{minipage}
\end{table*}

\section{Discussion}

\subsection{Confirmed Cepheids based on NIR observations}

For all confirmed OCs, our best-fitting DMs are consistent to within
0.1 mag with the corresponding determinations in the literature.

\paragraph{BB Sgr and Collinder 394}
BB Sgr was first confirmed as a member of Collinder 394 by
\citet{Turner85}. Its DM and reddening are $(m-M)_{0,V}=9.04 \pm 0.08$
mag and $E(B-V)=0.25 \pm0.01$ mag, respectively \citep{Turner85},
which is within 1--2$\sigma$ of our NIR determinations,
$(m-M)_{0,J}=9.35 \pm 0.16$ mag and $E(J-H)=0.095\pm0.005$ mag,
respectively. The BB Sgr--Collinder 394 combination meets all five
selection criteria.

\paragraph{WZ Sgr and Turner 2}
WZ Sgr was first confirmed as a member of Turner 2 by
\citet{Turner93}, who derived its DM and reddening,
$(m-M)_{0,V}=11.26$ mag and $E(B-V)=0.56$ mag, respectively, which
compare very well with our results, $(m-M)_{0,J}=11.22\pm0.11$ mag and
$E(J-H)=0.18 \pm 0.01$ mag. WZ Sgr and Turner 2 also satisfy all five
selection criteria.

\paragraph{V Cen and NGC 5662}
V Cen was first suggested as a member of NGC 5662 by
\citet{Turner82}. With $(m-M)_{0,V}=9.10 \pm 0.08$ mag and
$E(B-V)=0.31 \pm 0.01$ mag, its optical values are consistent with our
NIR determinations, $(m-M)_{0,J}=9.24 \pm 0.11$ mag and $E(J-H)=0.10
\pm 0.01$ mag, respectively. V Cen and NGC 5662 also satisfy all five
selection criteria.

\paragraph{CV Mon and van den Bergh 1}
The membership probability of CV Mon of the OC van den Bergh 1 was
studied in detail by \citet{Turner98b}. It is characterized by
$(m-M)_{0,V}=11.08 \pm 0.03$ mag and $E(B-V)=0.75 \pm 0.02$ mag, which
is again similar to our results, $(m-M)_{0,J}=11.11 \pm 0.21$ mag and
$E(J-H)=0.26 \pm 0.015$ mag. The CV Mon--van den Bergh 1 pair also
meets all five selection criteria.

\paragraph{RU Sct and Trumpler 35}
The membership probability of RU Sct of the OC Trumpler 35 was studied
in detail by \citet{Turner80}. We determined
$(m-M)_{0,J}=11.43\pm0.16$ mag and $E(J-H)=0.33 \pm 0.01$ mag, which
is consistent with $(m-M)_{0,V}=11.60 \pm 0.12$ mag and $E(B-V)=1.03
\pm 0.02$ mag from \citet{Turner80}. In the context of our
instability-strip selection, RU Sct is located at $1.8 \sigma$ from
the ridge line.

\paragraph{X Cyg and Ruprecht 175}
The X Cyg membership probability of Ruprecht 175 was studied in detail
by \citet{Turner98a}. We find $(m-M)_{0,J}=10.23\pm0.16$ mag and
$E(J-H)=0.065\pm0.01$ mag, which is marginally similar (given the
uncertainties quoted) to their determinations, $(m-M)_{0,V}=10.43 \pm
0.04$ mag and $E(B-V)=0.25 \pm 0.02$ mag, respectively. X Cyg and
Ruprecht 175 satisfy all five selection criteria.

\paragraph{S Nor and NGC 6087}
S Nor and NGC 6087 were studied in detail by \citet{Turner86}. Their
determinations, $(m-M)_{0,V}=9.78 \pm 0.03$ mag and $E(B-V)=0.19 \pm
0.12$ mag are again similar to our results, $(m-M)_{0,J}= 9.84 \pm
0.11$ mag and $E(J-H)=0.06 \pm 0.005$ mag. S Nor and NGC 6087 satisfy
all five selection criteria.

\paragraph{CF Cas and NGC 7790}
CF Cas and NGC 7790 were studied in detail by \citet{An07}. They found
$(m-M)_{0,V}=12.46 \pm 0.11$ mag and $E(B-V)=0.48 \pm 0.02$ mag,
compared with our determinations of $(m-M)_{0,J}= 12.58\pm0.31$ mag
and $E(J-H)=0.14\pm0.02$ mag, respectively. CF Cas and NGC 7790
satisfy all five selection criteria.

\paragraph{U Sgr and IC 4725}
U Sgr and IC 4725 were also studied in detail by \citet{An07}, who
determined $(m-M)_{0,V}=8.93 \pm 0.03$ mag and $E(B-V)=0.39 \pm 0.03$
mag. These values are consistent with our results (in particular the
DM), $(m-M)_{0,J}= 8.90\pm0.11$ mag and $E(J-H)=0.18\pm0.010$ mag. U
Sgr and IC 4725 also satisfy all five selection criteria.

\paragraph{SX Car and ASCC 61, S Mus and ASCC 69, UW Car and Collinder
220, V379 Cas and NGC 129} These four OC Cepheids were first found by
\citet{Anderson13}. Here, we confirm their membership of their
postulated host clusters. All of these Cepheids meet all five
selection criteria to be associated with their proposed host OCs. We
calculated the respective NIR DMs and colour excesses, which are all
consistent with the values found by \citet{Anderson13}. As regards the
other new OC Cepheid proposed by \citet{Anderson13}, ASAS
J182714$-$1507.1, its $V$-band DM and proper motion are identical to
those of its host cluster (see also Table 2). However, we do not have
access to its NIR photometry, so that we cannot more robustly confirm
this OC--Cepheid pair.

\subsection{New Cepheids}

\paragraph{Dolidze 34 and TY Sct, CN Sct}
TY Sct is located at approximately 4 arcmin from the centre of Dolidze
34, while CN Sct is found at about 10 arcmin from the cluster
centre. Dolidze 34 is a poorly studied OC. We obtained $E(J-H)=0.36
\pm 0.02$ mag, $\log(t \mbox{ yr}^{-1}) = 7.9 \pm 0.2$ and
$(m-M)_{0,J}=11.85 \pm 0.21$ mag. Except for the DM, these values are
close to those listed by \citet{Kronberger06}, $E(B-V)= 1.041$ mag,
$\log(t \mbox{ yr}^{-1}) = 7.95$ and $(m-M)_{0,V}=10.76$. The proper
motion of TY Sct is similar (to within $1 \sigma$) of that of Dolidze
34; the proper motion of CN Sct is found within $1.3 \sigma$ of the
cluster's mean proper motion. In addition, the age of CN Sct, $\log(t
\mbox{ yr}^{-1}) = 7.64 \pm 0.1$, is consistent with that of the host
OC, $\log(t \mbox{ yr}^{-1}) = 7.9 \pm 0.2$. CN Sct lies in the
cluster's instability strip, while TY Sct is located at $1.5 \sigma$
from the instability strip's central ridge line. \citet{Fernie95} list
radial velocities for TY Sct and CN Sct of 25.50 km s$^{-1}$ and 19.7
km s$^{-1}$, respectively, which are inconsistent with the radial
velocity of Dolidze 34. However, the cluster's radial velocity of 7 km
s$^{-1}$ \citep{Dias14} is based on a snapshot estimate using a single
star, which happens to be CN Sct. Since CN Sct is a Cepheid, this
determination of a `radial velocity' is fraught with uncertainty and
should therefore not be relied upon.

\paragraph{Dolidze 52 and XX Sgr}
Dolidze 52 is also poorly studied. We determine a best-fitting
reddening of $E(J-H)=0.24 \pm 0.015$ mag and a DM, $(m-M)_{0,J}=10.77
\pm 0.21$ mag. XX Sgr is located at approximately 15 arcmin from the
centre of Dolidze 52, while its proper motion is consistent with the
cluster's to within $1 \sigma$.

\paragraph{NGC 6683 and CK Sct}
CK Sct is located at about 20 arcmin from the centre of NGC 6683. Its
proper motion is similar to the average proper motion of NGC 6683. In
the cluster's CMD, CK Sct is found near the instability strip's ridge
line. We derive $E(J-H)=0.18 \pm 0.02$ mag, $\log(t \mbox{ yr}^{-1}) =
8.0 \pm 0.2$ and $(m-M)_{0,J}=11.62 \pm 0.21$ mag. DAML02 lists
$\log(t \mbox{ yr}^{-1}) = 7.0$ for NGC 6683 and an apparent diameter
of 3 arcmin. This discrepancy is the reason why this potential OC
Cepheid is usually excluded by other authors. However, based on our
CMD, NGC 6683 could be consistent with an older age, while the
apparent diameter of the cluster could be as large as 10 arcmin, which
is supported by the locus of the main-sequence ridge line of stars
located within 10 arcmin of the cluster centre. The radial velocity of
NGC 6683 is $v_{\rm r}=4.0\pm10.0$ km s$^{-1}$ \citep{Dias14} -- based
on only one star -- which is comparable with that of CK Sct $v_{\rm
  r}=-0.4$ km s$^{-1}$ (although the large error bar makes this
comparison rather meaningless).

\paragraph{Feinstein 1 and U Car}
U Car is found at $\sim 1^\circ$ from the centre of Feinstein 1, which
corresponds to approximately twice the cluster's size. The reddening,
age and DM derived here are similar to the equivalent data provided in
DAML02. The DMs of U Car -- $(m-M)_{0,V}=11.07\pm0.32$ mag -- and
Feinstein 1 -- $(m-M)_{0,J}=10.88\pm0.21$ mag -- are also similar; the
proper motion of U Car, $\mu_{\alpha,{\rm Cep}} = -4.8$ mas yr$^{-1}$,
$\mu_{\delta,{\rm Cep}} = 2.4$ mas yr$^{-1}$, is found within the
1$\sigma$ range of the cluster's central proper motion,
$\mu_{\alpha,{\rm cl}} = -3.80$ mas yr$^{-1}$, $\mu_{\delta,{\rm cl}}
= 2.15$ mas yr$^{-1}$ (adopting a radius of 8.7 arcmin).
\citet{Anderson13} obtained $\delta v_{\rm r}=-15.13$ km s$^{-1}
\equiv 1.6\sigma$; their cluster radial velocity was provided by
\citet{Kharchenko07} on the basis of one star only, which prompted
\citet{Anderson13} to state that this cluster needs more study. The
radial velocity of Feinstein 1 which we adopted is $v_{\rm
  r}=2.20\pm5.0$ km s$^{-1}$, as derived by \citet{Dias14} based on
four stars. This is, within the uncertainties, consistent with the
value for U Car, $v_{\rm r}=0.21$ km s$^{-1}$.

\paragraph{ASCC 61 and VY Car}
VY Car and SX Car are both located near the outer edge of the OC ASCC
61. Their proper-motion and DM measurements imply that they are robust
cluster members. \citet{Anderson13} suggested that SX Car may be an OC
Cepheid, while they considered VY Car's status inconclusive because
the latter Cepheid's age is found in the $2 \sigma$ range. We obtained
$\Delta \log(t \mbox{ yr}^{-1})= 0.5$, while VY Car is selected as a
cluster member on the basis of the proper-motion and DM measurements.

\subsection{Uncertain cluster Cepheid}

\paragraph{ASCC 64 and XZ Car}
XZ Car is located approximately 17 arcmin from the centre of ASCC 64,
which is somewhat outside the cluster radius: the cluster's diameter
is 21.6 arcmin \citep{Kharchenko05}. XZ Car meets the proper-motion,
age and instability-strip selection criteria. The radial velocities of
ASCC 64 and XZ Car are $v_{\rm r}=1.15\pm5.77$ km s$^{-1}$
\citep{Dias14} and 3.14 km s$^{-1}$, respectively. However, the
cluster's DM, $(m-M)_{0,J}=11.24\pm0.21$ mag, is approximately 1 mag
smaller than that of XZ Car, $(m-M)_{0,J}=12.26\pm0.10$ mag.
\citet{Anderson13} quote a difference in radial velocity of $\Delta
v_{\rm r}=-11.465$ km s$^{-1}$; their OC proper motion was taken from
\citet{Kharchenko07}, which is based on only one star. The difference
in parallax is $\Delta \varpi=0.305$ mas, which is more than twice as
large as $\sigma_{\varpi}=0.136$ mas, so that they consider this a
low-possibility OC Cepheid.

\subsection{Rejected cluster Cepheid}

\paragraph{NGC 4349 and R Cru}
R Cru is located at roughly 16 arcmin from the centre of NGC 4349,
which is well beyond the size of this cluster (5 arcmin diameter). The
age of NGC 4349 is $\log(t \mbox{ yr}^{-1}) = 8.6$, its reddening is
$E(J-H)=0.10\pm0.02$ mag and the best-fitting DM is
$(m-M)_{0,J}=11.04\pm0.25$ mag, values which are identical to those of
\citet{Majaess12b}. R Cru is located above the cluster's instability
strip, and its PLR DM is less than $9.8 \pm 0.1$ mag, which implies
that R Cru is a foreground Cepheid, located of order 1 kpc closer to
us than NGC 4349.

\subsection{Reddening and data limitations}

$V$-band mean magnitudes are available for most Cepheids. However, NIR
mean magnitudes are available for only a small subsample. This means
that we need to investigate the reddening affecting each Cepheid
instead of using the average reddening values towards their host
OCs. This is because at optical wavelengths, differential reddening
across some of our clusters may be significant, sometimes reaching
0.5--1.0 mag in the $V$ band. Such an uncertainty would seriously
affect our distance selection. In the NIR, this concern can be
alleviated, because the differential reddening would be reduced to at
most 0.2--0.3 mag, which is comparable with the typical error in our
DMs.

For most clusters, we use 2MASS data for our isochrone
fits. Unfortunately, some clusters exhibit unclear main sequences,
which particularly applies to some of the new Cepheids associated with
small OCs. In addition, some clusters' main sequences yield large
error bars in our isochrone fits, preventing us from confirming some
other OC Cepheids (e.g., Berkeley 58, Turner 9). However, for some
northern OCs, UKIDSS data is available, while for some southern OCs we
have access to VVV data. If these observations are added to 2MASS at
the faint end, the isochrone-fitting results are much
improved. \citet{Majaess11} combine 2MASS and (as yet proprietary) VVV
data, which results in a more obvious main sequence for the OC
Lyng{\aa} 6, and hence an improved membership assessment of TW Nor in
Lyng{\aa} 6. \citet{Majaess12b} use VVV data to undertake a detailed
study of NGC 4349; they robustly show that R Cru is not a member of
NGC 4349.

\subsection{Proper-motion and radial-velocity limitations}

Proper motions and radial velocities provide very tight constraints on
likely OC members. The PPMXL Catalogue provides reliable proper
motions for all stars in the regions occupied by our sample
OCs. However, it is sometimes difficult to exclude field stars from
our calculations of the clusters' average proper motions. In other
words, the errors in the clusters' proper motions are sometimes large
and thus may render the proper-motion constraints we employed
invalid. This difficulty is also linked to the photometric
uncertainties inherent to the data used.

Average radial velocities and their errors for our sample clusters are
listed in the DAML02 OC calalogue, originally based mostly on the
compilations of \citet{Dias14} and \citet{Kharchenko13}. However,
radial velocities have only been calculated for fewer than 1000 OCs,
and more than half of these radial velocities are based on fewer than
three stars. This clearly leads to large uncertainties in our
radial-velocity database.

\section{Conclusions}

We have analysed all potential combinations of Galactic Cepheids and
OCs in the most up-to-date catalogues available. Isochrone fitting and
proper-motion calculation were done for every potential OC--Cepheid
combination. Five selection criteria were used to select possible OC
Cepheids: (i) the Cepheid of interest must be located within 60 arcmin
of the OC's centre, (ii) the Cepheid's proper motion is located within
the $1\sigma$ distribution of that of its host OC, (iii) the Cepheid
is located in the instability strip of its postulated host OC, (iv)
the Cepheid and OC DMs should differ by less than 1 mag, and (v) the
Cepheid and OC ages should be comparable: $\Delta \log (t \mbox{
  yr}^{-1}) <0.3$. A comparison of the reddening values derived from
our isochrone fits with the reddening values of \citet{Tammann03}
shows that our results are reliable. The NIR PLR is obtained for a
confirmed sample of about 19 OC Cepheids, $M_J= (-3.12 \pm 0.29)
\log(P \mbox{ day}^{-1})-(2.17 \pm 0.29)$ mag, which is in excellent
agreement with the best NIR PLR available for all Galactic Cepheids
$M_J= (-3.030 \pm 0.022) \log(P \mbox{ day}^{-1})-(2.306 \pm 0.020)$
from \citet{Ngeow12}.

Nineteen possible OC Cepheids are found based on our NIR analysis;
eight additional OC--Cepheid associations may be genuine pairs for
which we lack NIR data. However, the distance accuracy of our
isochrone fits is limited by the data quality, reddening and
membership probabilities of the OCs. Six of these Cepheids are new,
high-probability OC Cepheids, since they lie on the NIR PLR. These
objects include TY Sct and CN Sct in Dolidze 34, XX Sgr in Dolidze 52,
CK Sct in NGC 6683, VY Car in ASCC 61, and U Car in Feinstein 1. Two
additional new OC Cepheids lack NIR data: V0520 Cyg in NGC 6991 and CS
Mon in Juchert 18. However, many new OC Cepheids are still waiting to
be found with the availability of better proper-motion data, large
samples of radial-velocity data and high-precision and large-area NIR
data. In addition, aided by {\sl Gaia}, trigonometric parallaxes of
many Galactic Cepheids will be obtained in the next few years, and
direct calibration of the PLR will be possible.

\appendix

\section{Graphical representations of the loci of the other
  open cluster Cepheids in our diagnostic diagrams}

\clearpage

\begin{figure*}
\includegraphics[width=140mm]{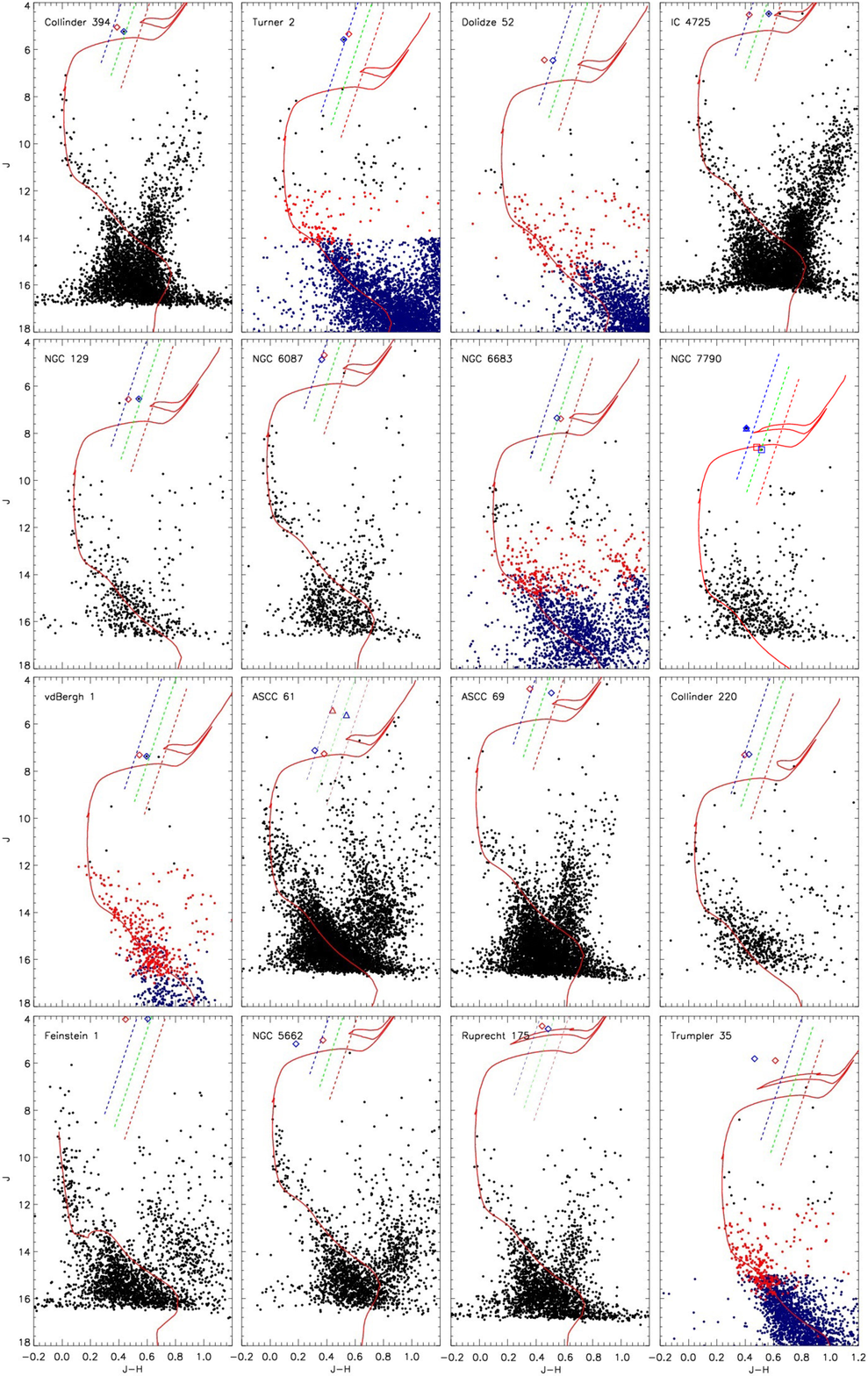}
\caption{CMDs of our sample OCs and the loci of their associated
  Cepheids.}
\end{figure*}

\clearpage

\begin{figure*}
\includegraphics[width=150mm]{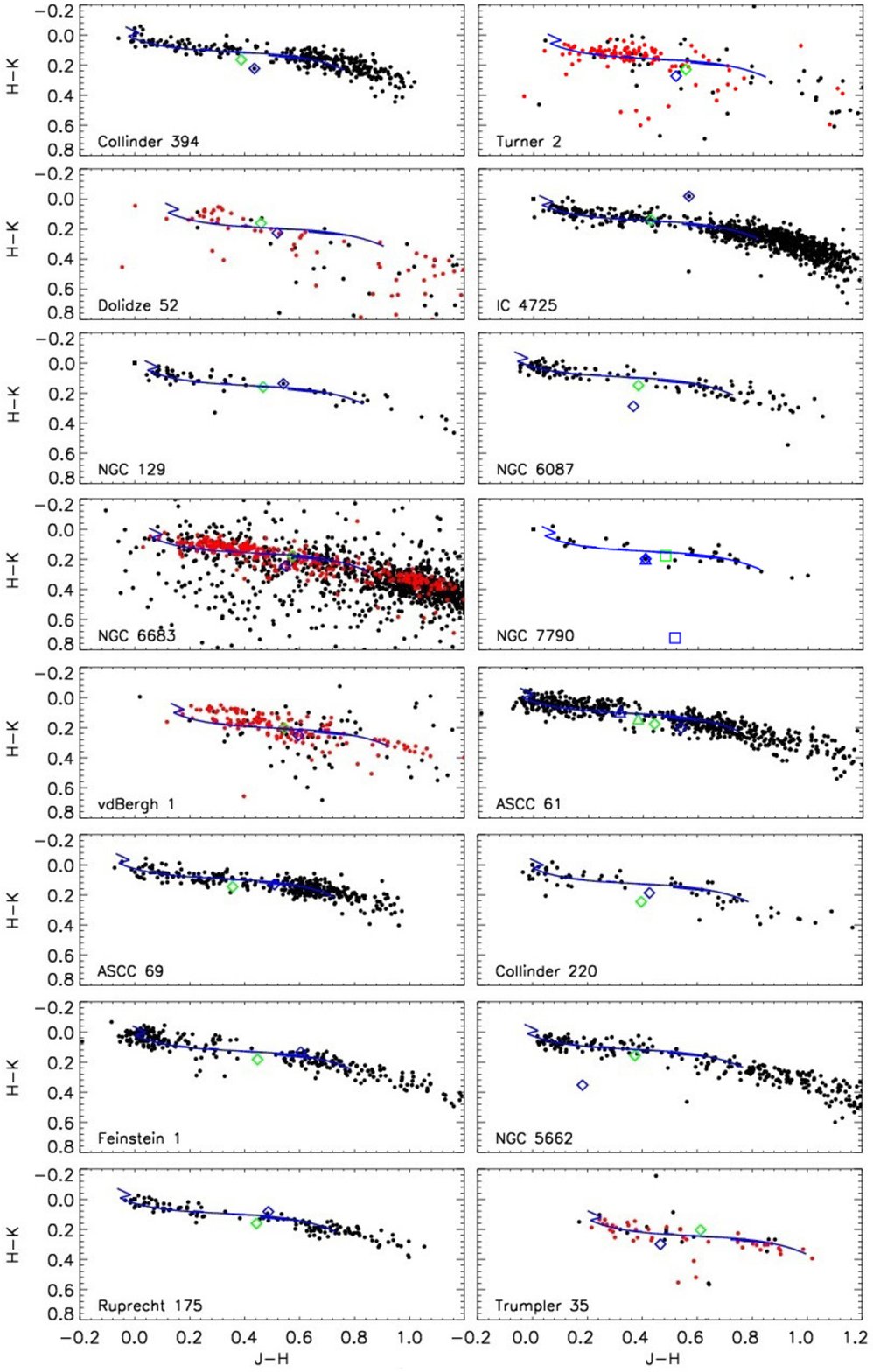}
\caption{Colour--colour diagrams of our sample OCs and the loci of
  their associated Cepheids.}
\end{figure*}

\clearpage

\begin{figure*}
\includegraphics[width=160mm]{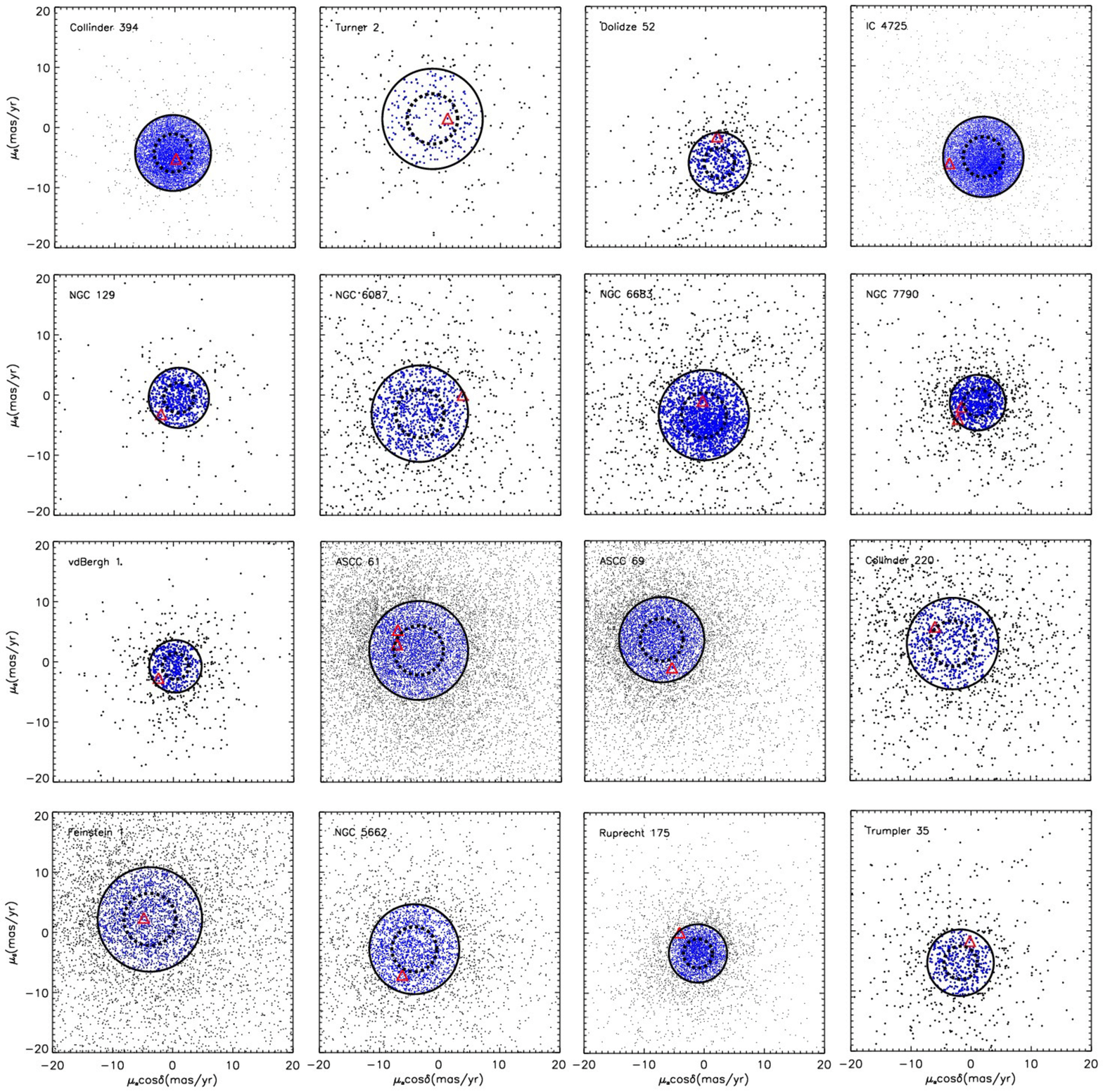}
\caption{Proper motion diagrams of our sample OCs and the loci of
  their associated Cepheids.}
\end{figure*}

\label{lastpage}

\end{document}